\def\ket#1{\left|\, #1\right\rangle}
\def\bra#1{\left\langle #1\right|}
\def\bracket#1#2{\left.\left\langle #1\right|#2\right\rangle}

\documentclass{article}
\usepackage{hyperref}
\usepackage{amsfonts}
\title{Feynman's Integral Is About Mutually Unbiased Bases}
\author{George Svetlichny\footnote{Departamento de Matem\'atica, Pontif\'{\i}cia Universidade Cat\'olica, Rio de Janeiro, Brazil \newline
svetlich@mat.puc-rio.br \hfill \url{http://www.mat.puc-rio.br/\~svetlich}}}
\begin{document}
\maketitle

\begin{abstract}
The Feynman integral can be seen as an attempt to relate, under certain circumstances, the
quantum-information-theoretic separateness of mutually unbiased bases
to causal proximity of the measuring processes.
\end{abstract}

\section{Introduction}

The Feynman integral, also know as the path integral, has achieved the status of a universal quantization procedure by which classical lagrangian theories are turned into quantum theories though a formally systematic, though in practice at times haphazard, procedure. In spite of its undeniable power, Feynman path integration has never achieved a well defined mathematical formulation. The Feynman ``integral'' does not deal with a conventional \(\sigma\)-additive measure and there are no general means to overcome this. We argue here that the Feynman integral expresses a scheme for interrelating quantum-information-theoretic systems such that measurement bases related by causality must  also  be related by being to a degree mutually unbiased.  The lack of a proper understanding of this apparently contradictory requirement is probably one of the reasons that the integral has never achieved a true mathematical status. Mutual unbiasedness is a relatively new topic, especially in the infinite-dimensional case, but is highly pertinent to understanding the Feynman integral. In this paper we shall present the connection between the Feynman integral and mutual unbiasedness, but unfortunately we'll still be far short of identifying the true mathematical nature of the object. Hopefully though, our remarks will contribute to this goal.

Nowadays there are many attempts to construct a more fundamental theory of physics, one that somehow combines quantum mechanics with general relativity, variously dubbed as ``theory of everything" or ``final theory" or, in its more modest version, ``quantum gravity" or ``quantum geometry". In all these approaches, the space-time continuum becomes an emergent effective object generated by some more fundamental substratum. In some approaches even quantum mechanics itself is seen as an emergent structure. In our view, given the great success of the Feynman integral method, a better understanding of what the integral is all about would help in the search for the elusive fundamental theory. If the  viewpoint of emergence is correct, then the Feynman integral itself is an emergent effective tool. A better understanding of it would help us see what it emerges from.

\section{The non-relativistic Schr\"odinger equation.}
Consider the non-relativistic Schr\"odinger's equation for a single particle of mass \(m\) in a time-independent potential \(V\). For simplicity we assume that space is one-dimensional.  Feynman's expression for the amplitude that a particle goes from position \(y\) at  time \(0\) to position \(x\) at time \(t\) is  an ``integral" over all paths \(s\mapsto x(s)\) with \(x(0)=y\) and \(x(t)=x\):

\begin{equation}\label{nrfeyint}
\bracket{x,t}{y,0}=\int \exp\left(\frac i\hbar\int_0^t L(x(s),\dot x(s))\,ds\right)\,{\cal D}x
\end{equation}

Here
\[L(x,\dot x)=\frac12 m\dot x^2-V(x)=\frac{p^2}{2m}-V(x)\]
is the lagrangian and the ``measure" \({\cal D}x\) is formally
\begin{equation}\label{Dx}{\cal D}x=\prod_s\,dx(s),\end{equation}
an ill-defined ``Lebesgue measure" over the infinite-dimensional space of possible particle positions for each time instance \(s\). In certain circumstances and with proper care, discrete approximations to the above integral do converge to the correct answer. The amplitude \(\bracket{x,0}{y,t}\) is then the kernel of the unitary time evolution operator \(U(t)\), meaning that the evolution of a wave function \(\psi(x,t)\) is given by
\begin{equation}\label{kernel}
\psi(x,t)=\int \bracket{x,t}{y,0}\psi(y,0)\,dy=\int K_V(x,y,t)\psi(y,0)\,dy
\end{equation}
where we have also introduced the notation \( K_V(x,y,t)\) for the kernel to indicate its dependence on the potential \(V\).

We begin our analysis of the Feynman integral by reversing the historical steps and rewrite the exponential of the action as a continuous product.

\begin{equation}\label{expILasprod}
e^{\frac i\hbar\int_0^t L(x(s),\dot x(s))\,ds}=\prod_0^t \left(e^{\frac i\hbar L(x(s),\dot x(s))}\right)^{ds}
\end{equation}

For those unfamiliar with the continuous product, its Riemann version is defined for a function \(f(t)\) on an interval \([a,b]\) in strict analogy with the Riemann integral as the limit of a product resulting from a partition of the interval:
\[
\prod_a^bf(t)^{dt} =
\lim_{N\to\infty}\prod_{i=1}^Nf(t_i')^{\Delta_it}
\]
(The Riemann integral would be the limit of the Riemann {\em  sum\/} \(\sum_{i=1}^Nf(t_i'){\Delta_it}\)). It's easy to see that if \(\ln f(t)\) is Riemann integrable then
\[
\prod_a^bf(t)^{dt} = \exp\left(\int_a^b \ln f(t)\, dt\right)
\]
and  in particular for { \(f(t) = \exp(g(t))\)}
\[
\prod_a^b\left(e^{g(t)}\right)^{dt} = \exp\left(\int_a^b g(t)
\,dt\right).
\]
Thus the expression (\ref{expILasprod}) is mathematically sound for continuous paths \(x(s)\). We now formally combine the product of the two products (\ref{expILasprod}) and (\ref{Dx}) into a single product and write the amplitude \(\bracket{x,t}{y,0}\) as
\begin{equation}\label{feyisiofprod}
\bracket{x,t}{y,0}=\int \cdots \int\prod_s \left(e^{\frac i\hbar L(x(s),\dot x(s))}\right)^{ds}\,dx(s).
\end{equation}

Though this looks like a integral of a product measure, it is not, as \(L\) depends on \(\dot x\) which involves the values of \(x(s)\) in an infinitesimal neighborhood of \(s\).
Historically it is basically (\ref{feyisiofprod}) that was deduced for the amplitude \(\bracket{x,t}{y,0}\). One chooses a set of intermediate times \(0=s_0<s_1<s_2<\cdots<s_N=t\) and inserts the complete position operator basis at times \(s_i\), formally, \(I=\int\ket{x,s_i}\bra{x,s_i}\,dx\). The amplitude \(\bracket{x,s_i}{y,s_{i-1}}\) is then under some assumptions estimated to be \cite{feyn-hibb}
 \begin{equation}\label{apfeyamp}\left(\frac{2\pi i\hbar \Delta_is}m\right)^{-1/2}e^{\frac i\hbar L(x,y)\Delta_is}
 \end{equation} where \(\Delta_is=s_i-s_{i-1}\) and
\begin{equation}\label{feyamp}
L(x,y)=\frac{m}2\left(\frac{x-y}{\Delta_is}\right)^2-V(x),
\end{equation}
where of course one now identifies the first term with \(\frac12mv^2=\frac{p^2}{2m}\), the kinetic energy. Expression (\ref{feyamp}) can also be obtained by the Trotter product formula for the unitary evolution
\[U(t)=e^{-\frac i\hbar Ht}=\lim_{n\to \infty}\left(e^{-\frac i\hbar H_0t/n}e^{-\frac i\hbar Vt/n}\right)^n\]
where \(H_0=\frac{p^2}{2m}\). For small \(t\) one can use \(n=1\) to get a good approximation. This can be calculated exactly and coincides with (\ref{apfeyamp}). Passing on now to more and more refined partitions, one arrives at (\ref{feyisiofprod}) in the ``continuum limit". The next historical step is to rewrite the product of phases as an exponential of an integral which is now identified as the action and arrive at (\ref{nrfeyint}). In our view, based on the idea of emergent space-time, form (\ref{feyisiofprod}) is more natural as it suggests that the phase \(\left(e^{\frac i\hbar L(x(s),\dot x(s))}\right)^{ds}\) is of local origin and arises through the emergence process along with the space-time structure. We'll come to this point again later.

It's instructive to look at the form of the amplitude \(\bracket{x,t}{y,0}\) after a {\em  finite\/} number of insertions of  complete position bases at intermediate times. Assuming these times are sufficiently close together that one can use (\ref{apfeyamp}), and assuming all \(\Delta_is\) are equal to \(t/N\) one has the approximation
\begin{equation}\label{disap}
\bracket{x,t}{y,0}\approx \left(\frac{2\pi i\hbar t}{Nm}\right)^{-N/2}\int\cdots \int \prod_{i=1}^{N-1} e^{\frac i\hbar L(x_i,x_{i-1})(t/N)}\,dx_i
\end{equation}
where  \(x_i\) indicates \(x(s_i)\) the position at time \(s_i\).

Now (\ref{disap}) is up to an overall numerical factor an integral of a product of phases, and one should ask what is the significance of the fact that is precisely phases that are being integrated. No get further insight into this, consider now a finite-dimensional Hilbert space of dimension \(M\) and an inner product \(\bracket\Phi\Psi\) of two vectors. As in the Schr\"odinger evolution case let us insert a number \(N\) of complete orthonormal bases \(\ket{j,i}\) where \(i=1,\dots,N\) labels the bases and \(j=1,\dots, M\) labels the elements of the \(i\)-th base for \(i\) fixed. For heuristic reasons we indicate the bases as \(\ket{x_i}\) where now \(x_i\) for fixed \(i\) takes on values from \(1\) to \(M\). Thus we have:

\begin{equation}\label{justip}\bracket\Phi\Psi=
\sum_{x_1=1}^M\cdots\sum_{x_N=1}^M\bracket\Phi{x_1}
\bracket{x_1}{x_2}\cdots\bracket{x_{N-1}}{x_N}\bracket{x_N}\Psi
\end{equation}

Writing the sums as integrals over discrete measures one can express this as:

\begin{equation}\label{jusipasi}
\bracket\Phi\Psi=\int\cdots\int\left(\bracket\Phi{x_1}\prod_{k=1}^{N-1}
\bracket{x_k}{x_{k+1}}\,dx_k\right)\bracket{x_N}\Psi\,dx_N
\end{equation}

Up to an overall factor expression (\ref{jusipasi}) would be identical to (\ref{disap}) if we had chosen \(\Phi\) and \(\Psi\) to be eigenvectors of bases \(\ket{x_1}\) and \(\ket{x_N}\) respectively. Now for (\ref{jusipasi}) to be, up to an overall factor and the two end amplitudes, an integral over a product of phases, it's necessary and sufficient that all the inner products \(\bracket{x_k}{x_{k+1}}\) have the same modulus, that is
\(|\bracket{x_k}{x_{k+1}}|=1/{\sqrt M},\)
in which case
{ \[\bracket{x_k}{x_{k+1}}=\frac1{\sqrt M}
\,e^{iL_k(x_k,x_{k+1})}\]}
 and so
\[\bracket\Phi\Psi=M^{-N/2}\int\cdots\int\left(\bracket\Phi{x_1}\prod_{k=1}^{n-1}
e^{iL_k(x_k,x_{k+1})}\,dx_k\right)\bracket{x_N}\Psi\,dx_N \]
in form identical to expression (\ref{disap}).

In quantum information theory two bases   \(e_i\), and \(f_j\) in a finite
dimensional Hilbert space of dimension  \(M\) are  called {\em  mutually unbiased\/} if
 \begin{equation}\label{mubs}|\bracket{e_a}{f_b}|=\frac1{\sqrt M}.\end{equation}
What this implies is that knowing the result of a measurement in the first basis gives no
information about possible subsequent measurements in the second
basis as all results are equally probable. We see now that the Feynman integral is about mutual unbiasedness, it's about potential destruction of information. This is what is implied by integration of products of phases.

A consequence of (\ref{mubs}) is that
\[\bracket{e_a}{f_b}=\frac{e^{iL(a,b)}}{\sqrt M}\]
where the ``lagrangian" \(L\) is constrained by the requirement of unitarity of
the matrix { \(\bracket{e_a}{f_b}\)}. Such a matrix without the \(1/\sqrt{M}\) factor is know as a (complex) Hadamard matrix, a topic still under active research. Under the idea of emergent space-time, if part of what emerges is a system of mutually unbiased bases, then lagrangians, or better yet, local actions, are true physical quantities as they would be the phases of inner products of eigen-elements from two such bases.

If the phase nature of the Feynman integrand is a prescription for mutually unbiased bases, then one should see this fact in the objects that Feynman's integral is supposed to produce. In the non-relativistic case this would be the unitary evolution group of the Schr\"odinger operator. The position space kernel of this group, otherwise known as the Feynman propagator, has to exhibit some form of mutual unbiasedness. We now examine this question, and in the next section consider the field-theoretic situation in the same light.

For the Shr\"odinger equation one deals with infinite-dimensional Hilbert spaces and the property of mutual unbiasedness and of the analog of Hadamard matrices in this context is a totally unexplored mathematical territory  (see Weigert and Wilkinson \cite{weig-wilk:arXiv0802.0394} for a recent treatment).  One very familiar example of two such bases is the position and momentum bases. One has \(\bracket xp=e^{ixp}\) and these bases are  mutually unbiased. Knowing exact position one know nothing about the momentum and vice-versa. Not so familiar are the position bases at two times for a free particle. Let \({\bf x}(t)\) be the Heisenberg position operator at time \(t\), that is \({\bf x}(t)=U(t){\bf x}(0)U(t)^*\) where \(U(t)\) is the unitary evolution operator (not necessarily for a free particle). For a free particle we have \cite{feyn-hibb}:
\begin{equation}\label{freeprop}\bracket{x,t}{y,0}=\left(\frac{2\pi i \hbar
t}{m}\right)^{-1/2}\exp\,\frac{im(x-y)^2}{2\hbar t}
\end{equation}
And we see that the position bases for different times for a free particle are
mutually unbiased. This can also be understood by the position-momentum uncertainty principle; exact knowledge of the position at time \(0\) means that momentum is totally undetermined and so the particle can at a subsequent time be anywhere with equal likelihood. The information-theoretic status of the prefactor \(({2\pi i \hbar
t}/{m})^{-1/2}\) in (\ref{freeprop}) is not clear. It makes of (\ref{freeprop}) the kernel of the unitary evolution group. With it ``fuzzy information", that is,  the position information contained in a square-integrable wave function, such as a gaussian, gets transformed continuously while exact position information is destroyed.  The Wick rotation of (\ref{freeprop}) is the heat kernel which besides describing heat diffusion also describes the evolution of the probability density of finding a particle undergoing Brownian motion. The heat kernel behaves very differently, it does not destroy exact position information, which would be an initial probability distribution given by a delta function. This initial condition spreads as a time-varying gaussian whose form is given precisely by the heat kernel. An initial gaussian for the Schr\"odinger equation also spreads, but the narrower we take the initial gaussian, the faster is the initial spread (as seen at a fixed small time \(t\) later, for instance), so as we approach an initial delta function, the spread approaches infinite speed and so exact position information is totally destroyed. One may wonder then how is it that fuzzy information is transformed continuously, for fuzzy information can be thought of as an ensemble of exact information. In the quantum mechanical case this is not entirely true as the initial wave-function can be complex and so could have a position varying phase, which is information beyond an ensemble of exact position information. Put another way, a wave function is a coherent mixture of exact position information, while an initial probability distribution of Brownian particles is an incoherent one. One must realize that though in (\ref{freeprop}) all information that would be given by the position basis is gone, the kernel does contain information concerning the initial position, which now is contained in the phase. This instantaneous transfer of exact position information to the phase is then the correct way of thinking about what we've called ``destruction of exact information".

For a particle in a potential, the bases \({\bf x}(0)\) and \({\bf x}(t)\) are generally not mutually unbiased. This is again heuristically understandable from the position-momentum uncertainty principle. After exact localization, the particle can have any momentum, but the presence of a potential will modify its ability to go anywhere and so the bases will in general not be mutually unbiased. As \(t\to 0\) though the two bases should become more and more mutually unbiased, a property that we shall call {\em  asymptotic mutual unbiasedness\/}. This property is plausible from the time-energy uncertainty relation. As \(t\) gets smaller the particle can ``borrow" more and more energy and so the potential becomes less and less relevant as to where the particle can be. One can also see this heuristically from the following calculation. Let { \(K_V(x,y,t)\)} be the kernel of { \(U(t)\)} for Schr\"odinger's equation
with potential { \(V\)}. One has for \(t>0\):
\begin{equation}\label{kvscale}
K_V(x,y,t)=\frac1{\sqrt t}K_{V_t}(x/\sqrt t,y/\sqrt t,1)
\end{equation}
where \(V_t(x)=t V(\sqrt t \,x)\). This follows by rescaling Schr\"odinger's equation according to \(t\mapsto st\), \(x\mapsto \sqrt s\,x\) and calculating the modified potential in the new equivalent differential equation.
Now as  { \(t\to0\)}, { \(V_t \to 0\)} so
\[K_V(x,y,\tau)\approx\frac
1{\sqrt\tau}K_0(x/\sqrt\tau,y/\sqrt\tau,1)=K_0(x,y,\tau).\]
This is not mathematically rigorous as we've not defined what it means for the two basis to ``become more and more mutually unbiased", but shows that we can expect this result once a proper definition is given.

A particularly instructive example, to which we'll refer later, is the kernel for the harmonic oscillator. Shr\"odinger's equation now is
\[i\hbar\dot \psi=-\frac{\hbar^2}{2m}\frac{d^2}{dx^2}\psi+\frac12m\omega^2x^2\psi,\]
and one has \cite{feyn-hibb}:
\begin{equation}\label{kharmosc}
\bracket{x,t}{y,0}=\sqrt{\frac{m\omega}{2\pi i\hbar\sin(\omega t)}}\exp\left\{
\frac{im\omega}{2\hbar\sin(\omega t)}\left((x^2+y^2)\cos(\omega t)-2xy\right) \right\}
\end{equation}
One sees that, in spite of the presence of the potential, the two position bases are mutually unbiased for all \(t\). As \(t\to 0\) one can set \(\cos(\omega t)\approx 1\) and \(\sin(\omega t)\approx \omega t\), and with this (\ref{kharmosc}) becomes the free particle propagator (\ref{freeprop}) in conformity with the observation that as the time intervals diminishes, the potential matters less and less. It is only for a restricted class of potentials that \({\bf x}(0)\) and \({\bf x}(t)\) are mutually unbiased for all \(t\). In one dimension \(V(x)\) must be polynomial and at most quadratic. To see this, if \(K_V(x,y,t)=N(t)e^{iA(x,y,t)}\) with \(N(t)>0\) and \(A(x,y,t)\) real (in the mutually unbiased case, it can always be put into this form), then ignoring \(y\) one would have a solution of Schr\"odinger's equation of the form \(N(t)e^{iA(x,t)}\). For simplicity assume units such that Schr\"odinger's equation takes the form \(i\dot \psi=-\psi_{xx}+V\psi\), then one has \({\dot N}/N=-A_{xx}\) and \(\dot A +A_x^2+V=0\).
From the first equation one deduces \(A(x,t)=\frac12R(t)x^2+S(t)x+P(t)\) with \(R=-\dot N/N\). Substituting into the second equation, one obtains \(\frac12\dot R(t)x^2+\dot S(t)x+\dot P(t)+R(t)^2x^2+2R(t)S(t)x+S(t)^2+V(x)=0\).
This equation can only be satisfied if the coefficients of  \(x^2\), \(x\) and \(1\) (ignoring V) are constants: \(\frac12\dot R+R^2= k_1\), \(\dot S+2RS= k_2\), and \(\dot P+S^2= k_3\) and so \(V(x)=-k_1x^2-k_2x-k_3\). All the equations can now be solved in terms of elementary functions, but we won't need the solutions here.  Not all solutions however lead to such kernels as one that does must in the limit of \(t\to 0\) approach, in the sense of distributions, a multiple of \(\delta (x-y)\) for some point \(y\).  For potentials  that are not of the above form the two position bases cannot be mutually unbiased for all \(t\) but must approach this as \(t\to 0\).

 Mutual unbiasedness in the finite dimensional case is a relation of separateness. Being mutually unbiased is to be as far apart as possible in a natural metric defined on the set of bases. In fact such a metric is induced by an inner product in a real Hilbert space, and being mutually unbiased is equivalent to being orthogonal \cite{beng:quant-ph/0610216}. How one should measure the separation of bases in infinite dimension is not clear but being mutually unbiased is intuitively a form of separateness. It is then remarkable that as the two bases become temporally closer (\(t\to 0\)) they become more and more distant in information theoretic terms. This seemingly contradictory requirement, imposed by the form of the Feynman integral, is undoubtedly one of the factors involved in the repeated failures of discovering the true mathematical identity of this object.

One mathematical problem which should shed much light upon what Feynman's integral is all about is to answer the following question: What unitary groups \(U(t)\) in \(L^2({\mathbb R}^n)\) have the property that the position bases \({\bf x}(0)\) and \({\bf x}(t)=U(t){\bf x}(0)U(t)^*\) are   asymptotically mutually unbiased? Such groups would arise from lagrangian theories since for small \(t\) the kernel of \(U(t)\) would be approximately of the form \(N(t)\exp\left(iA(x,y,t)\right)\) and what lagrangians really are is the phase information in mutually unbiased bases. So \(A\) in a proper limit would produce the lagrangian. Having answered this question one would know for which lagrangian theories the Feynman integral would converge in some well defined sense and give the right answer. In a sense Feynman integral's mission in life is to produce exactly these unitary groups. The problem can be posed without any reference to Feynman's integral and so does not depend on us knowing what to do with the integral. Knowing the solution to the problem would however help us understand how the integral is supposed to behave, and thus clarify its mathematical nature.

A unitary group for which asymptotic mutual unbiasedness is not true is the translation group (on the line say): \((U(t)f)(x)=f(x+t)\). Here \({\bf x}(t)={\bf x}(0)+tI\) and the kernel of the group is \(K(x,y,t)=\delta(x-y+t)\). Exact information is not destroyed but is simply shifted by \(t\) units to the right, so these bases are not asymptotically mutually unbiased. The lagrangian for this dynamics would be simply \(p\) and so the exponential of the action would be \(\exp\left(\frac i\hbar\int_0^tmv(s)\,ds\right)=\exp\left(\frac{im}\hbar(x-y)\right)\) and the Feynman integral, to the extent that one can say it is defined, would be \(N\exp\left(\frac{im}\hbar(x-y)\right)\) where \(N\) is a the ``renormalized" value of the Feynman integral of the constant function \(1\). Feynman's integral does not give the right answer for the kernel of this unitary group.

Suppose we knew nothing about the ordinary integral but a lot about electrostatic potential problems (a somewhat unlikely scenario). Given a charge distribution \(\rho(x)\) in a bounded region of space, we know that we can do a multipole expansion at infinity and that the monopole contribution to the electric potential is of the form \(Q/r\). From our knowledge of potential theory we can now conclude that \(Q\) is a positive linear functional of \(\rho\) and might even argue toward \(\sigma\)-additivity. We then express \(Q\) as \(\int \rho(x)\,dx\) merely as a notational convenience, and would discover the mathematical properties of the integral by purely physical reasoning knowing what it is that the integral is supposed to give us, the monopole coefficient. For the Feynman integral the situation is not so easy, we know what the integral is supposed to give us, transition probabilities, but in most cases we have only the Feynman integral itself to calculate them and cannot deduce its properties from knowing what it must calculate if we have no independent way to calculate the same things. However, we now know that the Feynman integral must also provide us with asymptotically mutually unbiased bases and this is a problem that can be approached without using the  integral and so the nature of the integral can now be explored by independent means.

\section{Field Theory}

The relativistic field-theoretic situation is more complex, though the basic observations still hold. In this section we assume units such that \(\hbar=c=1\). For simplicity we consider a single real scalar field. The partition function is given by the Feynman integral:
\begin{equation}\label{ftfeyni}
Z=\int e^{iS(\phi)}{\cal D}\phi,
\end{equation}
where
\begin{equation}\label{acint}
S(\phi)=\int{\cal L}(x,\phi,
\partial_\mu\phi)\,d^4x,
\end{equation}
Here \({\cal L}\) is the lagrangian density and we once again write the exponential of the integral as a continuous product
\begin{equation}\label{conprod}
e^{i\int{\cal L}(x,\phi,
\partial_\mu\phi)\,d^4x}=\prod_x\left(e^{i{\cal L}(x,\phi(x),
\partial_\mu\phi(x))}\right)^{dx_1dx_2dx_3dx_4}=\prod_xe^{i\hat{\cal L}(x)\,d^4x}
\end{equation}
where the third term is an abbreviated notation for the second. Once again, formally,
\begin{equation}\label{formmeas}
{\cal D}\phi=\prod_x\,d\phi(x)
\end{equation}
and we end up with an expression similar to (\ref{feyisiofprod})
\begin{equation}\label{feyisiofprodf}
Z=\int\prod_xe^{i\hat{\cal L}(x)\,d^4x}\,d\phi(x).
\end{equation}
Again, this may look like a product measure but is not as \(\hat{\cal L}(x)\) depends on the derivatives of the field, and so on its values in an infinitesimal neighborhood of the point \(x\). This fact allows the field to propagate in physical space-time.

We feel that (\ref{feyisiofprodf}) is the proper way to look at the Feynman integral from the emergence viewpoint. The lagrangian, or better still, the infinitesimal actions \(\hat{\cal L}(x)\,d^4x\) are local emergent quantities that give us the phases of asymptotically mutually unbiased bases. Although much has been written about emergence, and how this idea is supposed to lead us to a more fundamental theory, there seems to be no real consensus as to what has to emerge from what and how. Emergent space-time is a favorite idea, but it's probably more correct to consider the emergence of the whole kit and caboodle, space-time, its material content, and local phases known as ``lagrangians", after all this is what we have in our universe as the result. It may be mathematically inconsistent to produce only part of this content.  Mutual unbiasedness is a physical property, it determines how certain type of measurements behave. Lagrangian theories is a mystical belief that some function and its concomitant variational calculus treatment determines physical reality. Seeing the relation of lagrangians to mutual unbiasedness suggests that lagrangians are not some fundamental starting point but a final product of some substratum that gives us systems of observables. The ontological status of the lagrangian of the world is not clear, it's a condition on behavior of measurements depending on what measurement tools one uses, what physical fields are involved in the make-up of the observing device.The emergence of a system of observables is the emergence process suggested by  Feynman's integral. Systems of observables is not a new idea, the whole of local quantum mechanics \cite{haag} is founded upon this premise. The systems examined up to now usually already assume a classical space-time background and have to obey some physically motivated conditions such as Einstein locality. An extra ingredient which has not been considered up to now, is that those complete bases that are causally proximate (in an appropriate sense) should also be to a large extent (again in an appropriate sense) mutually unbiased. Causal nearness must be, paradoxically, related to quantum-information-theoretic separateness.

To see what this should mean for (\ref{feyisiofprodf}) one should interpret it as the result of introducing a ``continuum" of complete bases into some inner product. One often interprets \(Z\) as the inner product of the in and out vacuum, that is \(Z={}_{\small \rm out}\!{\bracket00}_{\small \rm in}\), though this is not the only way to look at it. Complete bases can be, at least formally, obtained by a foliation of space-time by space-like Cauchy hypersurfaces on each one of which one chooses the field strength basis. Let \(\hat \phi(x,\sigma) \) be the Heisenberg field operators where \(\sigma\) labels the space-like surface and \(x\) is a point in the surface labelled by \(\sigma\). Under an appropriate coordinate system one would have \(\hat \phi(x,\sigma) =U(\sigma,\tau)\hat \phi(x,\tau) U(\sigma,\tau)^*\) for a unitary ``evolution" groupoid \(U(\sigma,\tau)\). We are interested in bases \(\ket{\alpha,\sigma}\) with \(\alpha\) a real function on \(\sigma\) and which diagonalize all the \(\hat \phi(x,\sigma) \) simultaneously (sometimes know as the Schr\"odinger basis), that is:
\begin{equation}
\hat \phi(x,\sigma)\ket{\alpha,\sigma}=\alpha(x)\ket{\alpha,\sigma}.
\end{equation}
Introducing a ``continuum" of such bases for all the sheets of the foliation into the inner product \({}_{\small \rm out}\!{\bracket00}_{\small \rm in}\) would result in (\ref{feyisiofprodf}) as is explicitly done in any number of quantum field theory texts. The asymptotic mutual unbiasedness property would now be that the transition amplitude
\begin{equation}\label{transf}
\bracket{\alpha,\sigma}{\beta,\tau}
\end{equation}
would tend to a pure phase with a prefactor depending only on the surfaces \(\sigma\) and \(\tau\) as these approach each other. To calculate (\ref{transf}) one in principle would evaluate the Feynman integral (\ref{feyisiofprodf}) by integrating over all field configurations \(\phi\) between the two hypersurfaces for which \(\phi=\alpha \) on \(\sigma \) and  \(\phi=\beta \) on \(\tau \).

This in fact can be done in Minkowski space for a free massive relativistic real scalar field where the foliation is by space-like surfaces of constant time. The integral  can be explicitly, though formally, computed  by stationary phase methods \cite{oeck:PLB622.172}. One has:
\begin{equation}\label{ffmub}
\bracket{\alpha,t}{\beta,0}=
N\exp\left(i\int d^3x(\alpha,\beta)
\frac {\omega}{\sin \omega t}\left(\begin{array}{cc} \cos\omega t&-1 \\ -1&\cos\omega t
\end{array}\right)\left(\begin{array}{c }\alpha \\ \beta
\end{array}\right)\right)
\end{equation}
where \(N\) is a time-dependent normalizing prefactor (independent of \(\alpha\) and \(\beta\))  and \(\omega\) is the operator \(\sqrt{p^2+m^2}\) with \(p_i=i\partial/\partial x_i\). As expected, this shows the mutual unbiasedness of the two bases for all time intervals. The similarity of this expression to the harmonic oscillator kernel (\ref{kharmosc}) is evident, as could be expected. To some extent (\ref{ffmub}) begs the question of the mathematical nature of the Feynman integral. In principle the expression is the kernel of the unitary time evolution in the Shr\"odinger basis, that is given a state \(\Psi(\alpha,t)\) in this basis one should have, in analogy with (\ref{kernel}),
\begin{equation}\label{schker}
\Psi(\alpha,t) = \int\bracket{\alpha,t}{\beta,0}\Psi(\beta,0)\,{\cal D}\beta,
\end{equation}
where again \({\cal D}\beta=\prod_x\,d\beta(x)\) is the ill-defined ``Lebesgue measure" on the set of all possible field configurations. This apparent circularity however does not preclude investigation of mutual unbiasedness as an independent problem, and the integral in (\ref{schker}) can be expected to be less problematic than the Feynman integral itself.

As \(t\to 0\), if we write \(\sin \omega t\approx \omega t\) and \( \cos\omega t\approx I\) then (\ref{ffmub}) becomes
\begin{equation}\label{fstop}
\bracket{\alpha,t}{\beta,0}=
N\exp\left(\frac it\int d^3x\,(\alpha-\beta)^2\right)
\end{equation}
which should be compared to the free particle propagator (\ref{freeprop}).
This describes approximately how the field will behave immediately after a simultaneous measurement of the field strength at all points (a rather idealized possibility). The absence of any ``off-diagonal" terms relating two different points in space means the field does not propagate in physical space but otherwise  evolves in the field-strength space at each point as a free Schr\"odinger particle.
This is in analogy with the harmonic oscillator behaving as a free particle for very short time intervals. It might be curious to note that the emergent geometry of the \((\phi,t)\)-space at any point is Galilean and field strengths can change with arbitrary rate. There is no analog in (\ref{fstop}) of the mass \(m\) in the free particle case (\ref{freeprop}) as with our units the field has dimension \({\rm L}^{-1}\) whereas particle position has dimension \({\rm L}\) and so the integral in (\ref{freeprop}) needs another dimensional prefactor for the exponent to be dimensionless. The physical mass of the field does not appear in (\ref{fstop}).

The same behavior is to be expected from an interacting field also, again reasoning from uncertainty relations. Heuristically kernel (\ref{fstop}) corresponds to the non-relativistic lagrangian \(4\)-form \({\cal L}\,dt\,dx^3=\frac12\dot\phi^2\,dt\,d^3x\) which is invariant under the rescaling \(t\to st\), \(\phi\to \sqrt s\,\phi\). If one applies the same rescaling to the lagrangian \(4\)-form of a self-interacting  real relativistic scalar field, \([\frac12\dot\phi^2-\frac12(\nabla \phi)^2+\frac12m^2\phi^2+V(\phi)]\,dt\,d^3x\), one gets the \(4\)-form \begin{equation}\label{scaf}
\left[\frac12\dot\phi^2-\frac12s^2(\nabla \phi)^2+\frac12s^2m^2\phi^2+sV(\sqrt s\,\phi)\right]\,dt\,d^3x.
\end{equation}
This is the analog of (\ref{kvscale}) on the lagrangian level. As \(s\to 0\) only the first term survives and the theory approaches that of kernel (\ref{fstop}). This of course is much more heuristic than the corresponding argument given previously for the non-relativistic Schr\"odinger equation, but the basic principle is the same.

It's not clear what other mutual unbiasedness requirements are imposed by the Feynman integral beyond that of space-time foliations by space-like Cauchy hypersurfaces. Oekl \cite{oeck:PLB622.172} introduces the notion of states on time-like hypersurfaces for the same free field considered above, also in the Schr\"odinger representation. He obtains a mutually unbiased transition probability similar to (\ref{ffmub}) between field configuration on two such hypersurfaces. The precise physical meaning of this construction is not clear, but suggests that field theoretic mutual unbiasedness probably extends beyond what was here considered.

\section*{Acknowledgements}
This research was partially supported by the Conselho Nacional de Desenvolvimento Cient\'{\i}fico e Tecnol\'ogico (CNPq), and the Funda\c{c}\~ao de Amparo \`a Pesquisa do Estado do Rio de Janeiro  (FAPERJ).


\begin{thebibliography}{xx}
\bibitem{feyn-hibb} R.~P.~Feynman and A.~R.~Hibbs, {\sl Quantum Mechanics and Path Integrals}, McGraw-Hill, 1965.
\bibitem{weig-wilk:arXiv0802.0394}Stefan Weigert and Michael Wilkinson,  ``Mutually Unbiased Bases for Continuous Variables'', arXiv:0802.0394.
\bibitem{beng:quant-ph/0610216} Ingemar Bengtsson, ``Three ways to look at mutually unbiased bases",     arXiv:quant-ph/0610216v1.
\bibitem{haag} Rudolf Haag, {\sl Local Quantum Physic\/}, Springer Verlag, Berlin, 1992.
\bibitem{oeck:PLB622.172} Robert Oeckl, {\sl Physics Letters B\/} {\bf 622} 172 (2005).
\end{thebibliography}
\end{document}